# Perturbational treatment of spin-orbit coupling for generally applicable high-level multi-reference methods

Sebastian Mai[1], Thomas Müller[2,a)], Felix Plasser[3], Philipp Marquetand[1], Hans Lischka and Leticia González[1]

[1]Institute of Theoretical Chemistry, University of Vienna, Währinger Str. 17, 1090 Vienna, Austria
[2]Institute for Advanced Simulation, Jülich Supercomputing Centre, Forschungszentrum Jülich,
52425 Jülich, Germany
[3]Interdisciplinary Center for Scientific Computing, University of Heidelberg, Im Neuenheimer Feld 368,
69120 Heidelberg, Germany
[4]Department of Chemistry and Biochemistry, Texas Tech University, Lubbock, Texas 79409-1061, USA

**Abstract**

An efficient perturbational treatment of spin-orbit coupling within the framework of high-level multi-reference techniques has been implemented in the most recent version of the COLUMBUS quantum chemistry package, extending the existing fully variational two-component (2c) multi-reference configuration interaction singles and doubles (MRCISD) method. The proposed scheme follows related implementations of quasi-degenerate perturbation theory (QDPT) model space techniques. Our model space is built either from uncontracted, large-scale scalar relativistic MRCISD wavefunctions or based on the scalar-relativistic solutions of the linear-response-theory-based multi-configurational averaged quadratic coupled cluster method (LRT-MRAQCC). The latter approach allows for a consistent, approximatively size-consistent and size-extensive treatment of spin-orbit coupling. The approach is described in detail and compared to a number of related techniques. The inherent accuracy of the QDPT approach is validated by comparing cuts of the potential energy surfaces of acrolein and its S, Se, and Te analoga with the corresponding data obtained from matching fully variational spin-orbit MRCISD calculations. The conceptual availability of approximate analytic gradients with respect to geometrical displacements is an attractive feature of the 2c-QDPT-MRCISD and 2c-QDPT-LRT-MRAQCC methods for structure optimization and ab inito molecular dynamics simulations.

The importance of relativistic effects for accurate electronic structure calculations involving heavy atoms is well documented.[1–4] Although kinematical (scalar) relativistic effects primarily manifest in the core electron distribution, their impact on the chemically relevant valence electrons is substantial either by direct stabilization of the core-penetrating s and – to lesser extent – p or by indirect destabilization of the d and f valence orbitals due to more efficient screening of the nuclear charge.[5,6] In addition, it is important to include non-scalar interactions, with spin-orbit (SO) coupling being the most important one. SO coupling has a large impact on the electronic properties of heavy atoms.[7,8] And although the SO coupling matrix elements of organic molecules containing only light atoms are usually small (often below 100 cm$^{-1}$), SO coupling introduces a mechanism for intersystem crossing (ISC), allowing for population transfer between states of different spin multiplicity.[9,10] In some cases, ISC can compete with internal conversion on an ultrafast time scale, profoundly affecting photochemistry. A few examples can be found in Refs.
11–19. In odd-electron systems, SO coupling fundamentally affects the dimensionality and topology of the photochemically relevant crossing seam.[20–23] Also, dissociation energies cannot be predicted accurately without incorporation of SO coupling.[24]

The Dirac equation[25] is the exact wave equation for a single electron in the context of special relativity, and therefore it naturally incorporates scalar-relativistic effects and SO coupling. Its solutions are wavefunctions (bispinors) consisting of four components. Thus, methods incorporating the Dirac equation are termed 4-component (4c) methods. For many-electron systems, there exists no exact analogue to the Dirac equation. An approximately relativistic many-electron equation is the Dirac-Coulomb-Breit equation,[26,27] which superficially resembles the
Schrödinger equation with the one-electron part of the Hamiltonian replaced by Dirac operators and the two-electron part amended by a retardation term.[28] Removing the retardation term yields the Dirac-Coulomb equation. Both equations retain the 4c-wavefunctions from the Dirac equation and form the basis for methods usually termed 4c electronic structure methods. The energy spectrum of the Dirac as well as the Dirac-Breit equation displays a continuum at negative energies extending to $-\infty$, rendering these equations unsuitable to variational solution schemes.
Hence, practical implementations invoke the no-pair approximation[2,29,30] and implicitly project out the undesirable negative-continuum states, leading to a 4c electronic Hamiltonian bounded from below. Examples of implementations of 4c methods are based on Hartree-Fock (termed Dirac-Fock),[31,32] configuration interaction (CI),[33] multi-configurational self-consistent field (MCSCF),[31,34] coupled cluster (CC),[35–37] and density functional theory (DFT).[38,39] In practice, however, the 1- and N-particle basis set requirements render the highly accurate, generally applicable 4c multi-reference electron correlation methods (multi-reference configuration interaction (MRCI), MRCC) an option for few-atom molecules, only, while more approximate 4c techniques have a wider scope.

A number of transformation and elimination techniques aim at transforming the Hamiltonian such that only the upper portion of the energy spectrum is retained, thereby also reducing the number of components of the wavefunction to two (spinors). In wide-spread use are the Douglas-Kroll-Hess (DKH) Hamiltonian[2,40] and the zero/first-order regular approximations (ZORA/FORA).[41] The resulting methods using these techniques are termed 2-component (2c) electronic structure methods. 2c methods still retain spin-dependent terms and, in the limit of an infinite order unitary transformation (DKH) or regular approximation, exactly reproduce the positive energy spectrum of the Dirac-Breit Hamiltonian. There exist implementations of 2c SCF,[42] CI,[33,43] MCSCF,[44] and CC[45] as well as 2c-DFT for ground state energies and gradients[46] and 2c-TD-DFT for the calculation of excited states including SO coupling.[47–49]

Omitting the SO terms yields scalar-relativistic Hamiltonians (such as the usual scalar DKH approximation) with wavefunctions containing a single component, leading to the 1c methods. An economical alternative are relativistic effective core potentials (RECPs) with and without SO coupling terms. They constitute parametrized potentials which reproduce either experimental or computed data.[50] Scalar-relativistic electronic structure methods are simple extensions to non-relativistic quantum chemistry codes.

Since 1c methods are not able to describe SO coupling and 4c methods are computationally very expensive, 2c methods are of most interest for this work. These can be divided[29,30,51] into methods incorporating SO coupling already in the orbital optimization step—yielding a spinor basis instead of a conventional, spin-averaged molecular orbital (MO) basis—and methods working with a basis of real, spin-averaged MOs, introducing SO coupling during the correlation treatment. Although both variants are equivalent in the limit of full CI, the latter is more approximate for incomplete electron correlation treatments: in the MO based formalism spin polarisation (i.e., different spatial extent and angular distribution of $\alpha,\beta$ spinor pairs due to SO effects) is not present at the MO optimization stage and has to be recovered in the correlation step by including singly excited CSFs in the wavefunction expansion. However, a MO basis works well for first- and second row elements, transition metals, and f block elements.[4,51] While MO based methods process solely real-valued two-electron integrals, a spinor basis invariably implies complex-valued integrals except for the point groups $C_{2v}$, $D_2$ and supergroups thereof. Hence, in terms of computational efficiency, molecules of low symmetry favor MO based electron correlation methods, while otherwise both approaches may be expected to be roughly equivalent for equal-sized N-electron basis set expansions.

SO coupling primarily lifts the spin and orbital degeneracy of electronic states with open-shell character and thus may lead to a high density of low-lying electronic states. The complicated interplay between electron correlation, scalar relativistic effects, and SO coupling tends to increase the multi-reference character of the electronic states, so that conventional single-reference methods are usually not applicable. A popular framework for the description of SO coupling is the family of 2c-MRCI methods.

In this work, we extend the existing 2c-MRCISD (MRCI including single and double excitations) implementation of the Columbus quantum chemistry program package[52–54] by quasi-degenerate perturbation theory (QDPT). Within QDPT, the SO Hamiltonian is constructed in a small basis of 1c-MRCI wavefunctions (the model space) and diagonalized to yield the 2c-wavefunctions. It thus separates the correlation treatment from the SO treatment, making it an approximation to variational 2c-MRCI methods, but operates essentially at the cost of 1c-MRCI methods. Besides 2c-QDPT-MRCISD, we also describe its approximately size-extensive 2c-QDPT-LRT-MRAQCC variant based on LRT-MRAQCC.[55]

The accuracy of both 2c-QDPT approaches with respect to their variational 2c counterparts is validated by MRCISD calculations on acrolein and its S, Se, and Te analoga. Acrolein is small enough so that accurate multi-reference calculations can be easily conducted. Additionally, acrolein features close-lying excited singlet and triplet states, so that the influence of SO coupling on the excitation energies and the shape of relevant potential energy curves (PECs) can be studied. The substitution of oxygen with its homologues allows to increase the strength of SO coupling while keeping the electronic structure otherwise mostly unaltered.

A. Hamiltonian and integrals

The Hamiltonian considered in this work can be expressed within second quantization[56,57] as

$$\hat{\mathcal{H}} = \sum_{ij} h_{ij} \hat{E}_{ij} + \frac{1}{2} \sum_{ijkl} [ij;kl](\hat{E}_{ij}\hat{E}_{kl} - \delta_{kj}\hat{E}_{il})$$
$$+ \sum_{ij\sigma\tau} h_{i\sigma,j\tau} \hat{E}_{i\sigma,j\tau}, \qquad (1)$$

where $\hat{E}_{ij}$ denotes spin-averaged excitation operators. The integrals $h_{ij}$ and $[ij;kl]$ are the spin-independent one- and two-electron integrals, respectively. The $\overline{h}_{i\sigma,j\tau}$ are spin-dependent (spin indices $\sigma$ and $\tau$) one-electron integrals describing the SO interaction. The SO integrals can be factored into an orbital and a spin-dependent part

$$\overline{h}_{i\sigma,j\tau} = \sum_{\gamma} (-1)^{\gamma} \langle i|q_{\gamma}|j\rangle \langle \sigma|s_{\gamma}|\tau\rangle, \qquad (2)$$

where orbital (qγ) and spin part (sγ) are expanded in terms of spherical tensors.[57] Scalar-relativistic effects are accounted for via appropriate one-electron integrals hij. One popular approach for the calculation of the hij are energy-[50,58] and shape-consistent relativistic pseudo-potentials[59,60] where scalar-relativistic effects are approximated by j-averaged (total angular momentum averaged) potentials. Other approaches are simplified relativistic treatments such as ab initio model potentials (AIMPs)[61] and the model core potentials (MCPs).[62] For all-electron calculations, the necessary integrals are available from scalar DKH theory.[2,63] Even though the SO operator originating from the Breit-Pauli formalism[26–28] is a one- and two-electron operator, the one-electron terms dominates.[4] Thus, one usually resorts to effective one- electron operators for the description of SO effects. Among the most popular implementations of these effective one-electron operators is the Atomic Mean Field Integral (AMFI) method.[64–67] Alternatively, relativistic pseudo-potentials can be extended by a j-dependent term representing the difference from the (scalar) j-averaged potentials,[50] making the SO integrals hiσ,jτ also available for calculations with pseudo-potentials.

## B. MRCISD

Retaining most of the 1c-MRCISD formalism in calculations at the 2c-MRCISD level of theory is an important asset as it allows to re-use many parts of the existing 1c machinery. In particular, the choice of a real, spin-averaged molecular orbital basis and a real Hamiltonian is beneficial for efficiency of the computationally expensive matrix-vector product step of the direct CI formalism.[68] A spinor-based implementation of various HF and post-HF methods is, e.g., offered by the DIRAC package.[69] In the standard 1c-MRCISD formalism, the wavefunction is linearly expanded in a basis $\{\Phi_i\}$ of spin-adapted configuration state functions (CSFs)

$$\Psi(\Gamma, S, M_s = S)^{1c} = \sum_{i=1}^{N_{CSF}} c_i \Phi_i(\Gamma, S, M_s = S). \quad (3)$$

The coefficients $c_i$ are determined variationally to minimize the total energy. The CSF basis $\{\Phi_i(\Gamma, S, M_s)\}$ is characterized by the irreducible representation (irrep) $\Gamma$ and the spin quantum numbers S and $M_s$. In absence of SO coupling, the Hamiltonian commutes with $\hat{S}^2, \hat{S}_z$ and spatial symmetry operators and thus in the basis $\{\Phi_i(\Gamma, S, M_s)\}$ the Hamiltonian assumes a block-diagonal form. Blocks sharing the same $(\Gamma, S)$ combination are identical irrespective of the $M_s$ value, so that it is sufficient to compute only solutions for blocks spanned by the CSF basis $\{\Phi_i(\Gamma, S, M_s = S)\}$.

The 2c-MRCISD wavefunction is equally expanded in a CSF basis. However, the one-electron SO operator introduces off-diagonal blocks in the Hamiltonian, according to general selection rule $|\Delta M_s| \leq 1$ and $|\Delta S| \leq 1$ (except for $S = S' = 0$). Thus, the CSF basis employed in a 2c calculation is the union of multiple $\{\Phi(\Gamma, S, M_s)\}$ expansion spaces, including all Ms components. The 2c- MRCISD wavefunction is given by

$$\Psi(\Gamma)^{2c} = \sum_{i=1}^{N_{CSF}} c_i \Phi_i(\Gamma, S, M_s)$$
$$\forall S \leq S_{max} \text{ and } |M_s| \leq S, \qquad (4)$$

where the expansion is truncated at some suitably chosen maximum spin multiplicity Smax. Again, the coefficients are obtained variationally, and SO coupling and electron correlation are treated on the same footing. Because we cannot omit the MS ≠ S blocks in the Hamiltonian, the 2c expansion is by a factor f 2c longer than the sum of their 1c constituents

$$f^{2c} \approx \frac{\sum_{S=S_{min}}^{S_{max}} N_{CSF}(S) \cdot (2S+1)}{\sum_{S=S_{min}}^{S_{max}} N_{CSF}(S)}, \qquad (5)$$

where NCSF(S) denotes the length of the 1c CI expansion with spin S. Assuming that the number of iterations per root is the same and that the computational effort per iteration scales approximately linear with the size of the CSF expansion, in absence of symmetry the 2c-MRCISD calculation is approximately f 2c times as expensive per root as all the 1c constituents together. An additional factor of 2 is required for the odd-electron case (see Sec. III B).

This type of variational 2c-MRCI calculation based on spin-averaged MOs (denoted "2c-MRCI" from here on) has been described in the literature, using different notions. Vallet et al.[70] refers to the method as "double group CI" (DGCI), while Kleinschmidt et al.[71] and Buenker et al.[72] refer to it as MRSOCI. Among the first variational 2c-CI treatments is the one proposed by Christiansen,[73] implemented in the CIDBG code.[74] Two recent implementations of the 2c-MRCISD method are available in the Columbus[57] and Spockci[71] programs. Both codes are based on an expansion in a CSF basis (in opposition to an expansion in a determinant basis). However, Spockci is a selecting CI code,[75] which minimizes the size of the configuration space by selecting individual CSFs for variational treatment based on their MP2 estimate to the total correlation energy (plus an optional specific treatment of single excitations), whereas in Columbus all single and double excitations with respect to any reference CSF are included. The latter approach facilitates the calculation of gradients, non-adiabatic couplings, as well as efficient vectorization.33,76

**C. QDPT**

Instead of simultaneously treating electron correlation and SO coupling, one commonly resorts to an a posteriori perturbative treatment of SO coupling. To this end, an initial set of 1c-MRCI states $\Phi_i(\Gamma, S, M_s)$ is used to expand the 2c-wavefunction

$$\Psi(\Gamma)^{2c-QDPT} = \sum_{i=1}^{N_{state}} \tilde{c}_i \tilde{\Phi}_i(\Gamma, S, M_s)$$

$$\forall S \leq S_{max} \text{ and } |M_s| \leq S_{max}. \quad (6)$$

Variational optimization of the coefficients $\tilde{c}_i$ defines a first-order QDPT treatment. Here, electron correlation and SO coupling effects are not treated on the same footing, since SO relaxation effects can only be described within the model space. Owing to the reduced flexibility compared to that of the fully variational scheme, accurate results require suitable size and adequate choice of the model space $\{\Phi_i(\Gamma, S, M_s)\}$.

Like the 2c-MRCISD method (see above), the 2c-QDPT-MRCISD scheme has been described by a number of authors with a quite diverse vocabulary (see Ref. 70 and references therein). For example, this method is frequently termed "two-step" method[77] or "interacting states."[24] Vallet et al.[70] refers to it by "CILS+SO." Since this approach may also be described as a MRCI in a contracted CSF basis—with the contraction coefficients coming from the 1c-MRCISD—Buenker et al.[72] denotes the method as "LSC-SO-CI" (LS-contracted SO-CI). The SPOCKCI developers[71] and the ORCA developers[78] refer to the approach as QDPT, which is the abbreviation we adopt here as well.

In contrast to the QDPT scheme in Spock-ci or Orca, the newly implemented Columbus 2c-QDPT-MRCISD method can make use of general reference spaces producing large scale – even multi-billion – CSF spaces without ever resorting to the underlying determinant expansion in order to evaluate the SO matrix elements. In fact, the time required to evaluate the SO matrix elements in the model space basis is almost negligible.

Since an effective Hamiltonian is constructed in the QDPT scheme, there are additional variations of this approach: SO-RASPT2[24,79] (SO-restricted active space second-order perturbation theory) expands the model space Hamiltonian at the RASSCF (restricted active space self-consistent field) level of theory and incorporates corrections for state-specific (SS) dynamic electron correlation effects derived from 1c-CASPT2 calculations. As mentioned above, approaches relying on a 1c-contracted model space usually face difficulties with spin-polarisation effects in small model spaces. The EPCISO method[77] (effective and polarized configuration interaction with spin-orbit) addresses this problem by projecting the contracted large CI expansion upon a much smaller uncontracted CI expansion (<105 determinants) used for the construction of the effective Hamiltonian in the second step including corrections for electron correlation from the initial step. Here, the major issue is the potential for imbalanced treatment of electron correlation and SO coupling. Molpro accounts for SO coupling80 in a fashion similar to the 2c-QDPT-MRCISD scheme of Columbus, albeit with the model space expanded in even more rigid internally-contracted MRCI wavefunctions and without the possibility to obtain fully variational 2c-wavefunctions.

### D. MRAQCC and LRT-MRAQCC

By construction, any truncated CI method is neither size-consistent nor size-extensive. While a posteriori Davidson-type techniques (MRCISD+Q) can be applied in order to obtain size- extensivity corrections to the total energy, there is no corresponding wavefunction available. Hence, the SO coupling matrix elements cannot be consistently evaluated, preventing a combination of QDPT and MRCISD+Q. Replacing the diagonal matrix elements of the model space Hamiltonian by the MRCISD+Q energies while computing the off-diagonal SO coupling matrix elements from the MRCISD wavefunctions is a rather heuristic approach and lacks consistency.

LRT-MRAQCC,55 a perturbative extension to the state-selective MRAQCC method,81 offers a natural way to consistently derive approximately size-extensive ground and excited state energies as well as transition densities. Hence, diagonal and off-diagonal matrix elements of the model space Hamiltonian can be described consistently. Both LRT-MRAQCC and MRAQCC belong to a family of correlation energy functionals, that can be cast into an MRCISD eigenvalue problem with diagonal CI matrix elements shifted by $\Delta_0$,82

$$\langle \Phi_i | (\mathcal{H} - E_0^\alpha) + \underbrace{\sum_{k \notin \text{int}} (1-G)\Delta E_0 | \Phi_k \rangle \langle \Phi_k |}_{\Delta_0} \left( \sum_j c_j^\alpha \Phi_j \right) = \Delta E^\alpha c_i^\alpha. \quad (7)$$

The projection operator $\sum_{k \notin \text{int}} |\Phi_k\rangle\langle\Phi_k|$ ensures that solely matrix elements of non-internal CSFs (CSFs with electrons in external orbitals or with excitations out of the internal orbitals doubly occupied in all reference CSFs) are modified. The method-specific constant G equals 1 for MRCISD and 1-(ne −3)(ne −2)/(ne-1)ne for MRAQCC, with ne being the number of correlated electrons. In case of MRAQCC, $\Delta E_0 = \Delta E\alpha$ is the correlation energy computed with the MRAQCC functional with respect to the energy of the reference wavefunction ($\alpha$). Since $\Delta E\alpha$ occurs on both sides of the equation, it is computed iteratively by reinserting the current estimate until convergence. In case of LRT-MRAQCC, ΔE0 is the MRAQCC correlation energy of the reference state (usually the ground state), i.e., LRT-MRAQCC is no longer a state-specific functional and the electronic states are mutually orthogonal as they share the same Hamiltonian.

Compared to MRCISD, MRAQCC and LRT-MRAQCC impose stricter requirements on the construction of the reference space: MRAQCC assumes that the reference wavefunction is a qualitatively good description of the correlated wavefunction while LRT-MRAQCC imposes the somewhat weaker constraint that the reference space is sufficiently flexible, as to describe all states of interest. For (higher) excited states, both methods are known to suffer potentially from intruder states. While the reference (ground) state energy coincides with the MRAQCC energy, the excited states do not and the error is related to the extent the MRAQCC correlation energies are similar for ground and excited states. Increasing the size of the reference space is a remedy to these discrepancies.

### III. IMPLEMENTATION

COLUMBUS 7.052, 54 contains a direct,68 efficiently parallelized76 implementation of non-relativistic and 1c-MRCISD as well as variational 2c-MRCISD.57 The necessary scalar-relativistic and SO integrals are delivered either by the ARGOS integral code,83 which is based on the pseudopotential approach, or by the SEWARD code from the MOLCAS package.84 SEWARD is able to provide the necessary integrals within the scalar-DKH and AMFI frameworks.

In order to extend the 2c-MRCISD machinery to the cases of 2c-QDPT-MRCI and 2c-QDPT-LRT-MRAQCC the key element is a fast algorithm to translate a given electronic state $\Phi_i$ ($\Gamma$, $S$, $s = S$) from the 1c to the 2c representation, thereby adding the $M_s$ components omitted in the 1c calculation. The algorithm must also be able to properly handle symmetry. Once the model space basis has been completely constructed, the existing 2c-MRCISD code57 can be used to evaluate the one-electron SO coupling elements

$$\langle \tilde{\Phi}_i(\Gamma, S, M_s) | \hat{\mathcal{H}}^{SO} | \tilde{\Phi}_{i'}(\Gamma', S', M'_s) \rangle \qquad (8)$$

without resorting internally to a determinant expansion. As the algorithm is closely tied to the representation of 1c and 2c wavefunctions within the Graphical Unitary Group Approach (GUGA),56 relevant aspects of GUGA and symmetry are briefly discussed below, before algorithm and program workflow are presented.

### A. GUGA representation

The tensor product space $(2n)^{\otimes N}$ spanned by N electrons in 2n spin orbitals forms a basis for the unitary group U(2n). For a spin-independent Hamiltonian, the wavefunction can be factorized into an orbital and a spin part and correspondingly U(2n) ⊃ U(n) ⊗ U(2).

Choosing a basis of spin-adapted CSFs ensures this factorization. According to the Pauli principle, only the basis of the totally antisymmetric irrep of U(2n) is admissible for fermionic wavefunctions, which for any chosen S, N uniquely fixes the irreps of U(n) and U(2). For a spin-independent Hamiltonian, a single member ($M_s = S$) out of the (2S+1) dimensional basis of U(2) is sufficient. Hence, the basis of U(n) suffices to enumerate the CSF basis. The generators of the unitary group satisfy the same commutation relations as the spin-averaged excitation operators (see Eq. (1)), so that the group theoretical apparatus of the unitary group allows to use U(n) to uniquely enumerate the CSF basis and to efficiently evaluate the matrix elements of the Hamiltonian.

The great achievement of Shavitt[56] was to translate the algebraic representation of U(n) into a compact graph (Shavitt graph) along with a small set of rules to evaluate the Hamiltonian matrix elements directly and efficiently from the graph without referring to complicated and lengthy algebraic expressions. The GUGA[56] paved the road to rather compact and fast implementations of a general MR-CISD code capable of running multi-billion MR-CISD calculations[76] due to extensive use of recursion, dense linear algebra, and parallelization techniques.

The Shavitt graph (Fig. 1) consists of a collection of enumerated vertices organized in an array structure connected by arcs. The levels j are associated with the spatial orbitals $\phi_j$ while the columns indicate cumulative spin coupling in terms of a and b (N = 2a + b, S = 2b). Each CSF is represented as a directed walk from tail to head and characterized by the unique sequence of $a_i, b_i, i = 1…n$ values associated with the vertices visited. The four different slopes of the arcs indicate the change in cumulative spin coupling and number of electrons due to the orbital associated with the vertex the arc connects to. Each partial graph consisting of all possible directed walks from tail to some vertex i at level j is a basis for U(j) constrained to $N_i$ electrons and spin $S_i$, which reflects that the basis of U(n) can be uniquely labelled by the chain U(n) ⊃ U(n − 1) ⊃ … ⊃ U(1). In the spin-independent case, there is only a single head, since the entire CSF space is characterized by $\{N, S, M_s = S\}$. In contrast, the spin-dependent CSF space is characterized by $\{N, S \leq S_{max}, -S \leq M_s \leq S\}$ giving rise to multiple heads (one per spin multiplicity). Each head additionally carries 2S + 1 extensions which represent the 2S + 1 components of the multiplet.

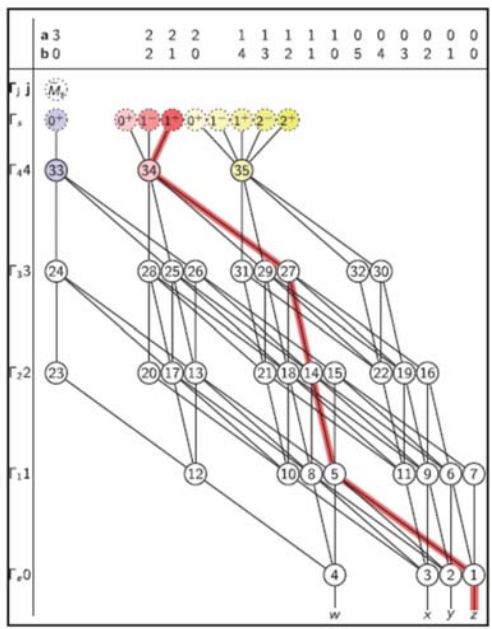

FIG. 1. Multi-headed Shavitt graph (internal part only, Smax= 2) for six electrons and four internal orbitals incorporating singlets (blue), triplets (red), and quintets (yellow). The red line illustrates a particular triplet CSF with associated symmetry-adapted spinfunction $|1, 1+\rangle$. For details, see text.

If all single and double excitations into the external orbitals are included, the graph has a particularly simple and regular structure for the external orbital levels. Thus, only the internal orbitals are explicitly included plus four interface vertices at level $j = 0$ connected to zero (z), one (y), two triplet-coupled (x), and two singlet-coupled electrons (w) in the external orbital space. While this approach favors efficiency by use of dense linear algebra, it rules out any individual selection of CSFs with a non-zero occupation of the external orbitals. Rather the CSF space encoded in a given graph can be manipulated by deleting arcs or vertices or by marking individual internal paths as invalid.

The CSF basis is enumerated continuously traversing the graph from the leftmost to the rightmost internal paths. Since the 1c graphs are subgraphs of the 2c graph, the relative ordering of a pair of CSFs is preserved within each spin multiplicity in both 1c and 2c representations. Enumerating the CSF basis in terms of the basis of U(n) renders the CI vector coefficients independent of Ms, so that all $2S + 1$ degenerate components are trivially accessible from the CI coefficients of its 1c counterpart of a given spin multiplicity. Hence, the translation of a 1c wavefunction expansion into its 2c counterpart can be accomplished by a simple restructuring of the CI vector elements. Figure 2(a) gives an example of this restructuring.

**B. Symmetry treatment, spin functions, and the odd electron case**

Columbus is restricted to Abelian point groups with one-dimensional irreps in order to simplify point group symmetry treatment. The symmetry of any CSF in the spin-independent case is evaluated as the direct product of the irrep associated with each arc passed in the internal walk

$$\Gamma^{1c} = \Gamma_e \otimes (\Gamma_1 \otimes \Gamma_2 \ldots \otimes \Gamma_n). \qquad (9)$$

$\Gamma_e$ denotes the contribution of the external orbital space, while $\Gamma_j$ is the symmetry of the orbital associated with internal level j if singly occupied. For unoccupied and doubly occupied orbitals, the factor is totally symmetric.

Since GUGA relies only implicitly on the spin subgroup U(2), in 1c calculations the spin functions are not fully specified and any unitary transformation of the 2S + 1 standard spin functions $|S, M_s\rangle$ is admissible. In the spindependent case, a symmetry-adapted basis of spin functions is chosen which transforms as the cartesian components of the angular momentum. Additionally, the phase is chosen such as to ensure real Hamiltonian matrix elements for the bosonic even-electron case.[57] This basis is denoted $|S, \tilde{M}\rangle$ with $\tilde{M}$ = 0+, . . . S−, S+ Hence, the symmetry treatment of the bosonic case solely requires the additional factor s representing the symmetry of the spin function (2c, even = 1c ⊗ s).

In absence of an external magnetic field, the Hamiltonian also commutes with the time-reversal symmetry operator. As a consequence, in the fermionic odd-electron case all electronic states are 2-fold degenerate (Kramer's degeneracy) and the degenerate components form a Kramer's pair.[85] They transform as the (one- or two-dimensional) fermionic irreps of the respective double group and the matrix elements of the SO contribution to the Hamiltonian are in general complex. The presence of multi-dimensional irreps and complex linear algebra renders the odd-electron case not directly suitable for the GUGA approach.

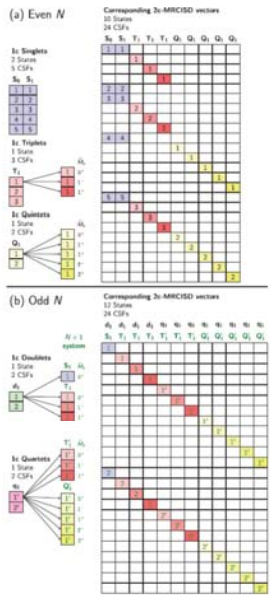

FIG. 2. Sketch of the step of translating the 1c-MRCI vectors into the model space components in 2c format (even electron case in (a); odd electron case (b)). The 1c-MRCI vectors of S0, S1, T1, and Q1 are shown on the left. The numbers in boxes indicate the enumeration of the CSFs in the 1c case while the labels right of the boxes label the color coding of the associated spin functions (symmetry adapted bases, see text) for the 2c case. The synthetic $N + 1$ electron system (see Sec. III B) is denoted in green text. Capital letters (S0, T1,... ) denote even-electron states, lowercase letters (d1, q1) denote oddelectron states. The 2c model space vectors are shown on the right. Empty boxes represent vanishing coefficients.

Thus, a single, fictitious, non-interacting electron residing in a fictitious, totally symmetric orbital is added to the odd-electron system yielding a synthetic (N + 1) even-electron system with a CSF space of twice the size.57 This is equivalent to replacing the Hermitian eigenvalue problem by the corresponding real eigenvalue problem of twice the dimension which yields a pairwise doubly degenerate eigenvalue spectrum. The latter reflects the arbitrary phase $e^{i\varphi}$ (with real $\varphi$) of the solutions to the initial Hermitian eigenvalue problem. With

$$\mathbf{H} = \mathbf{B} + i\mathbf{C}, \quad \mathbf{B} = \mathbf{B}^T, \quad \mathbf{C} = -\mathbf{C}^T$$

We have

$$\begin{bmatrix} \mathbf{B} & -\mathbf{C} \\ \mathbf{C} & \mathbf{B} \end{bmatrix} \left\{ \begin{bmatrix} \vec{u}_i \\ \vec{v}_i \end{bmatrix}, \begin{bmatrix} \vec{v}_i \\ -\vec{u}_i \end{bmatrix} \right\} = \lambda_i \left\{ \begin{bmatrix} \vec{u}_i \\ \vec{v}_i \end{bmatrix}, \begin{bmatrix} \vec{v}_i \\ -\vec{u}_i \end{bmatrix} \right\}. \tag{10}$$

The pairwise solutions are related to the initial Hermitian eigenvalue problem by $z_i = e^{i\varphi}(u_i + iv_i)$. Hence, the eigenvalue spectrum of the synthetic (N + 1) even electron problem is fourfold degenerate with only one-dimensional irreps. The four solutions are related to each other in a Kramer's basis through time-reversal symmetry and the subsequent mapping to the real eigenvalue problem. For the real Abelian double groups $D^*_2$, $D^*_{2h}$ and $C^*_{2v}$ each fermion irrep decomposes into four different irreps of the (N + 1) electron system. For $C^*_{2h}$, $C^*_s$, $C^*_2$ and $C_i$, $C_1$ there are two and one irreps, respectively. Double groups with inversion centre ($D^*_{2h}$, $C^*_i$, $C^*_{2h}$) have fermion irreps of even and odd parity, and correspondingly there are two sets of (N + 1) electron irreps of even and odd parity.

The multi-headed Shavitt graph for an odd-electron case (MR-CISD/CAS(5,4)) is displayed in Fig. 3. The additional fictitious non-interacting electron is represented in level $j = 5$. At level $j = 4$ the initial odd-electron system with its doublets (vertex #30) and quartets (vertex #31) is encoded coupled to the additional single electron: singlets and triplets arise from the doublets, while triplets and quintets arise from the quartets. Each of these vertices is complemented by the (2S + 1) symmetry-adapted spin functions of the synthetic N + 1 electron system. Figure 2(b) gives an example of the CI vector translation step for the odd-electron case.

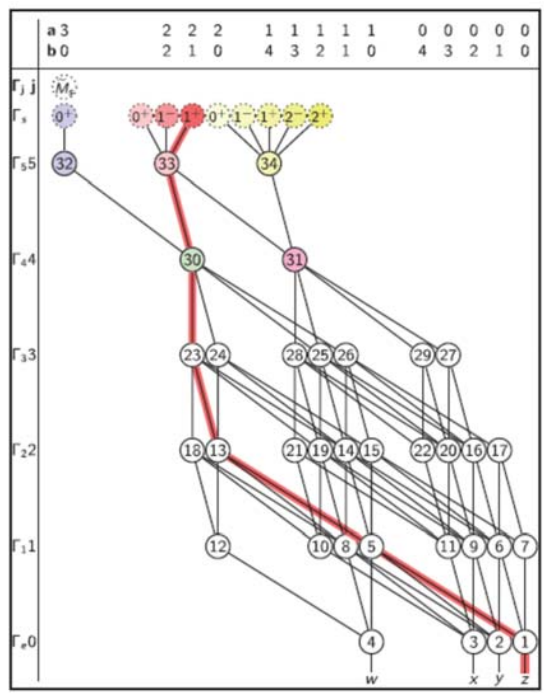

FIG. 3. Multi-headed Shavitt graph (internal part, only) for five electrons and four internal orbitals including doublets (green, vertex #30) and quartets (purple, vertex #31). Adding a non-interacting fictitious electron to an additional fictitious orbital (level j = 5) recovers the (N + 1) even electron formalism of twice the size of the N electron problem (singlets (blue), triplets (red), quintets (yellow)). The red line exemplifies a doublet CSF mapped to the |1, 1+⟩ component of a triplet CSF in the synthetic (N + 1) electron system. For details see text.

### C. QDPT

The program flow for a 2c-QDPT calculation is schematically depicted in Figure 4. The initial steps are the integral evaluation and the optimization of the spin-averaged MOs by SS or state- averaged (SA) MCSCF. Next, based on the chosen $S_{max}$, the total symmetry of the 2c states of interest ($\Gamma_{2c}$) and the size of the model space we determine the number of independent 1c-MRCISD calculations, each specified by spin multiplicity S, spatial symmetry $\Gamma = \Gamma_S \otimes \Gamma_{2c}$, and the required number of roots. After the model space basis is optimized optionally in parallel, the 1c CI vectors are translated to the 2c representation and written to the restart file. Finally, the 2c-MRCISD program reads the wavefunctions from the restart file and—during the first iteration—solely evaluates the off-diagonal SO contributions to the model space Hamiltonian with the diagonal elements set to the total energy of the model space components. Diagonalizing this matrix yields the first-order QDPT estimate for the effect of SO coupling along with the eigenvectors in the basis of the 1c-MRCISD solutions with symmetry-adapted spin functions $|S, M\rangle$. Continuing the execution of the 2c-MRCISD code beyond the first iteration, the entire Hamiltonian is evaluated and on convergence the fully variational 2c-MRCISD result is obtained. Since the 2c-QDPT wavefunctions are good initial guesses, the variational 2c-MRCISD calculation can be expected to converge quickly.

In case of the approximately size-extensive 2c-QDPT-LRT-MRAQCC variant, as an additional step after the MO optimization step, we need to fix the reference state (which is usually the well- separated ground state, but not necessarily so) and to compute the 1c-MRAQCC energy for this state, in order to evaluate $\Delta 0$ in Eq. (7). Afterwards, the same procedure as for 2c-QDPT- MRCISD is followed with the sole exception that the 1c-MRCISD calculations are replaced by 1c-LRT-MRAQCC calculations using the previously determined diagonal shift. Upon assembly of the restart file, a 2c-LRT-MRAQCC calculation is run for a single iteration and diagonalizing the resulting subspace matrix yields the 2c-QDPT-LRT-MRAQCC estimate for the effect of SO coupling.

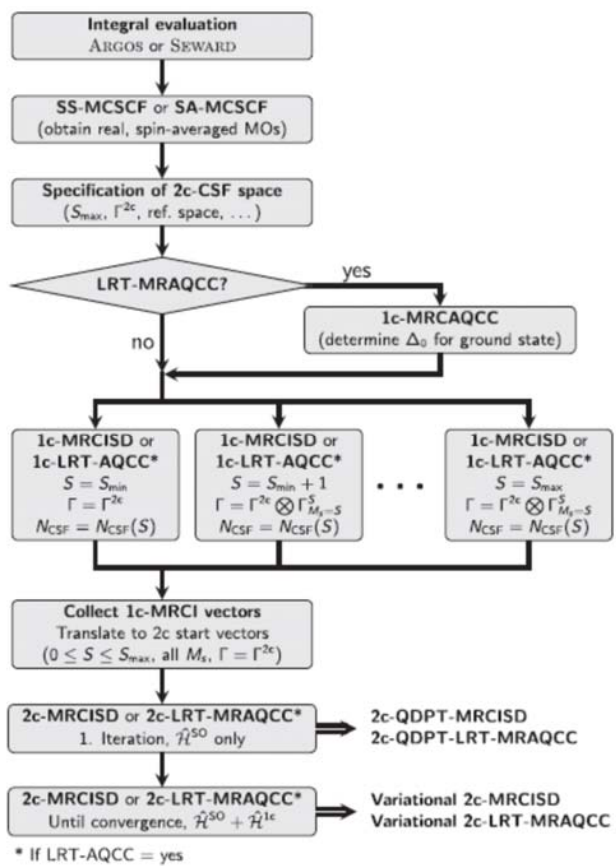

FIG. 4. Flow chart for the organization of 2c-QDPT-MRCISD, 2c-QDPTLRT-MRAQCC, and variational 2c-MRCISD based on initial start vectors taken from converged scalar-relativistic MRCISD or LRT-MRAQCC calculations.

## IV. APPLICATION

In order to compare 2c-QDPT-MRCI and variational 2c-MRCI (and their LRT-MRAQCC variants) for systems covering SO couplings from a few cm−1 up to more than 1000 cm−1, acrolein and its chalcogene analoga—with oxygen replaced by its homologues (S, Se, Te)—have been chosen. In the following, these systems are denoted O-acrolein, S-acrolein, Se-acrolein, and Te- acrolein. For simplicity, only the trans configurations of each of the molecules is included in the study. The geometry of acrolein with an arbitrary chalcogene Y is depicted in Figure 5. The objective of the calculations is to validate the QDPT ansatz with respect to the variational 2c-MRCI for the calculation of SO couplings in polyatomic molecules.

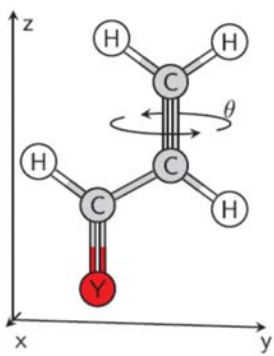

FIG. 5. Generic geometry of Y-acrolein and the C=C torsion coordinate θ.

The state-specific quality of the model space expansion in the QDPT calculations can be judged from the energy differences with respect to the fully variationally computed 2c states or by computing the overlap Oi of the model space wavefunctions with the fully variational wavefunction state $\Psi_i^{2c}$ of state i

$$O_i = \left\langle \Psi_i^{2c}(\Gamma) \middle| \sum_j \tilde{\Phi}_j(\Gamma, S, \tilde{M}) \right\rangle. \quad (11)$$

In the case of large spin-polarisation effects, high-energy singly excited CSFs will significantly contribute to the fully variational wavefunctions, but not to the QDPT wavefunctions. Thus, the deviation of the overlaps Oi from the ideal value of 1 will indicate that the model space is inadequate to correctly describe the 2c wavefunctions. Similarly, using the overlap Uij of the model space components {Φi } with the 2c and 2c-QDPT wavefunctions is an indicator of the extent of SO-coupling-induced mixing of the states

$$U_{ij}^{\text{even}} = \langle \Psi_i^{2c}(\Gamma) | \tilde{\Phi}_j(\Gamma, S, \tilde{M}) \rangle. \quad (12)$$

For the following calculations, the MRCI correlation treatment is based on real, spin-averaged MOs from (1c) CASSCF(6,5)/ANO-RCC-VDZP calculations, where state-averaging included the four lowest singlet and four lowest triplet states, except noted otherwise. Scalar-relativistic effects were treated with the DKH (second order) transformation technique. The reference space in the MRCI calculations is equal to the CAS space, including the nY orbital of the chalcogene and the π and π* orbitals. Inner shells were kept frozen. Single and double excitations are included as given below. All calculations were conducted without the explicit usage of symmetry.

### A. MRCISD vertical excitation energies

The S0 equilibrium geometries of the four systems were optimized with scalar-relativistic 1c-MRCISD. Table I presents the optimized geometry parameters. At these geometries, the energies of the 8 lowest 2c-states (approximately 2 singlets and 2 triplets with 3 components each) were calculated with the 2c-MRCISD and 2c-QDPT-MRCISD methods. In the latter case, the model space included 4 singlet and 4 triplet states. For comparison, the energy of the 2 lowest singlet and 2 lowest triplet states were also calculated using scalar-relativistic 1c-MRCISD. The energies relative to S0 and the weight of the dominant 1c wavefunction (according to Uij in Eq. (12)) are given in Table II.

TABLE I. Optimized S0 equilibrium geometries of Y-acroleins at the 1c-MRCISD level of theory (in Å and degrees).

| Y  | r(Y=C) | r(C–C) | r(C=C) | α(Y=C–C) | α(C–C=C) |
|----|--------|--------|--------|----------|----------|
| O  | 1.215  | 1.467  | 1.341  | 123.6    | 121.4    |
| S  | 1.636  | 1.449  | 1.347  | 124.9    | 122.2    |
| Se | 1.776  | 1.442  | 1.350  | 125.0    | 122.4    |
| Te | 1.982  | 1.433  | 1.352  | 125.2    | 122.8    |

TABLE II. Energies E (in eV) and main contribution Ueven (contr.) of the 1c-wavefunctions of the ground state and nπ∗ singlet and nπ∗ and ππ∗ triplet states of Y-acrolein. In the excitation energies, digits differing from the 2c-MRCISD result are bold.

| | 1c-MRCISD | | 2c-QDPT-MRCISD | | 2c-MRCISD | |
|---|---|---|---|---|---|---|
| State | contr. | $\Delta E$ | contr. | $\Delta E$ | contr. | $\Delta E$ |
| —Y=Oxygen— | | | | | | |
| 1 | $S_0$ | 0.0000 | 1.00 $S_0$ | 0.0000 | 1.00 $S_0$ | 0.0000 |
| 2 | $T_1$ | 3.4978 | 1.00 $T_1$ | 3.4977 | 0.99 $T_1$ | 3.4977 |
| 3 | $T_1$ | 3.4978 | 1.00 $T_1$ | 3.4977 | 0.98 $T_1$ | 3.4977 |
| 4 | $T_1$ | 3.4978 | 1.00 $T_1$ | 3.4978 | 1.00 $T_1$ | 3.4978 |
| 5 | $S_1$ | 3.7685 | 1.00 $S_1$ | 3.7683 | 1.00 $S_1$ | 3.7683 |
| 6 | $T_2$ | 3.8902 | 1.00 $T_2$ | 3.8902 | 0.99 $T_2$ | 3.8903 |
| 7 | $T_2$ | 3.8902 | 1.00 $T_2$ | 3.8902 | 0.99 $T_2$ | 3.8903 |
| 8 | $T_2$ | 3.8902 | 1.00 $T_2$ | 3.8904 | 1.00 $T_2$ | 3.8904 |
| —Y=Sulphur— | | | | | | |
| 1 | $S_0$ | 0.0000 | 1.00 $S_0$ | 0.0000 | 1.00 $S_0$ | 0.0000 |
| 2 | $T_1$ | 1.9002 | 1.00 $T_1$ | 1.9000 | 1.00 $T_1$ | 1.9000 |
| 3 | $T_1$ | 1.9002 | 1.00 $T_1$ | 1.9000 | 0.99 $T_1$ | 1.9000 |
| 4 | $T_1$ | 1.9002 | 1.00 $T_1$ | 1.9006 | 0.99 $T_1$ | 1.9006 |
| 5 | $S_1$ | 2.0897 | 1.00 $S_1$ | 2.0892 | 1.00 $S_1$ | 2.0892 |
| 6 | $T_2$ | 2.5716 | 0.99 $T_2$ | 2.5723 | 0.99 $T_2$ | 2.5724 |
| 7 | $T_2$ | 2.5716 | 1.00 $T_2$ | 2.5723 | 1.00 $T_2$ | 2.5724 |
| 8 | $T_2$ | 2.5716 | 0.99 $T_2$ | 2.5725 | 0.99 $T_2$ | 2.5726 |
| —Y=Selenium— | | | | | | |
| 1 | $S_0$ | 0.0000 | 1.00 $S_0$ | 0.0000 | 1.00 $S_0$ | 0.0000 |
| 2 | $T_1$ | 1.5693 | 0.97 $T_1$ | 1.5609 | 0.99 $T_1$ | 1.5630 |
| 3 | $T_1$ | 1.5693 | 0.98 $T_1$ | 1.5609 | 0.97 $T_1$ | 1.5632 |
| 4 | $T_1$ | 1.5693 | 0.99 $T_1$ | 1.5800 | 0.99 $T_1$ | 1.5810 |
| 5 | $S_1$ | 1.7228 | 0.95 $S_1$ | 1.7080 | 0.95 $S_1$ | 1.7080 |
| 6 | $T_2$ | 2.1067 | 0.98 $T_2$ | 2.1255 | 0.97 $T_2$ | 2.1279 |
| 7 | $T_2$ | 2.1067 | 0.98 $T_2$ | 2.1255 | 0.98 $T_2$ | 2.1280 |
| 8 | $T_2$ | 2.1067 | 0.95 $T_2$ | 2.1319 | 0.95 $T_2$ | 2.1341 |
| —Y=Tellurium— | | | | | | |
| 1 | $S_0$ | 0.0000 | 0.97 $S_0$ | 0.0000 | 0.97 $S_0$ | 0.0000 |
| 2 | $T_1$ | 1.1659 | 0.84 $T_1$ | 1.1121 | 0.84 $T_1$ | 1.1304 |
| 3 | $T_1$ | 1.1659 | 0.84 $T_1$ | 1.1121 | 0.83 $T_1$ | 1.1320 |
| 4 | $S_1$ | 1.2779 | 0.76 $S_1$ | 1.1988 | 0.78 $S_1$ | 1.2036 |
| 5 | $T_1$ | 1.1659 | 0.97 $T_1$ | 1.2323 | 0.97 $T_1$ | 1.2464 |
| 6 | $T_2$ | 1.5299 | 0.84 $T_2$ | 1.6497 | 0.84 $T_2$ | 1.6699 |
| 7 | $T_2$ | 1.5299 | 0.84 $T_2$ | 1.6497 | 0.84 $T_2$ | 1.6702 |
| 8 | $T_2$ | 1.5299 | 0.76 $T_2$ | 1.6759 | 0.78 $T_2$ | 1.6890 |

For O-acrolein, the S1 excitation energy of 3.768 eV fits nicely with the experimental value of 3.77 eV.86 Thus, we can tentatively assume that MRCISD gives an accurate description of the electronic correlation in the acrolein system. Since the valence shells of the heavier Y-acroleins are very similar to the one of O-acrolein, electronic correlation should be described well also for the heavier homologues.

For the four systems, the excited states S1, T1, and T2 are all close to each other in terms of vertical excitation energies. Thus, we can expect these states to mix notably upon inclusion of SO coupling. This also means that these states constitute the minimum model space for a reasonable QDPT treatment.

Most importantly, Table II compares the excitation energies from 1c-, 2c-QDPT-, and 2c-MRCISD. Since the fully variational 2c-MRCISD calculations includes electron correlation and SO coupling at the same level, it serves as the reference for the two other methods. For the light chalcogenes O and S, it can be clearly seen that 1c and 2c-QDPT treatments yield essentially the same excitation energies as the 2c-MRCISD calculation. In both cases, all excitation energies are within 1 meV of the reference energies. For the heavier Se-acrolein, errors with respect to the 2c results are still small, with the 1c calculation giving a maximum deviation of 27 meV for the highest state. 2c-QDPT shows a maximum deviation of only 2 meV. For the Te-acrolein, 1c- MRCISD is rather poor, giving errors in the excitation energies of up to 150 meV. 2c-QDPT performs better in this situation, with a maximum error of 20 meV.

The main contributions of the 1c states to the 2c-QDPT and 2c states are nice indicators of the extent of SO-induced state mixing. As clearly seen in Table II, in the O- and S-acrolein the states are very well described by a single 1c wavefunction. This of course correlates with the observation that for these systems also the excitation energies do not notably change when going from 1c to 2c methods. For Se- and especially Te-acrolein, the 2c states cannot easily be described by a single 1c wavefunction. In the Te-acrolein, there is also a notable splitting of the Ms components of the triplet states. This leads to the observation that the state with predominant S1 character lies in between the components of the T1 triplet. This can be explained by the strong interaction of S1 with T2—shifting down S1— and the interaction of S0 with T1, shifting up one component of T1. Naturally, this situation is not described by the 1c-MRCISD calculation, but it is quite well reproduced by the 2c-QDPT calculation. The observation that 2c-QDPT describes the 2c-MRCISD wavefunctions well can be explained with the overlaps of the 2c-wavefunctions with the model space. The amplitude of the overlaps $O_i$ are even in the case of Te-acrolein all above 0.99. Obviously, the model space captures most of the constituent CSFs of the 2c-wavefunctions. This in turn shows that spin-polarisation is not very important in the acrolein systems, as opposed to other cases reported in the literature.[24,72]

Figure 6 shows the mean absolute errors of the 1c and the 2c-QDPT excitation energies (Table II, columns 3 and 5) compared to the variational 2c-MRCISD (column 7). Note the logarithmic scale in Figure 6, which was chosen since the size of the SO couplings and hence the errors span several orders of magnitude when going from O-acrolein to Te-acrolein. It can be seen that 2c-QDPT in all cases gives better excitation energies than 1c calculations, reducing the mean absolute error by a factor of 5 or better. However, for systems with small SO coupling (O and S), even 1c-MRCISD is correct to within 1 meV. For the Te-acrolein, the 1c calculation is already off by 100 meV on average, which leads to a state ordering different than on the 2c level and may have a notable effect in, e.g., dynamical calculations.

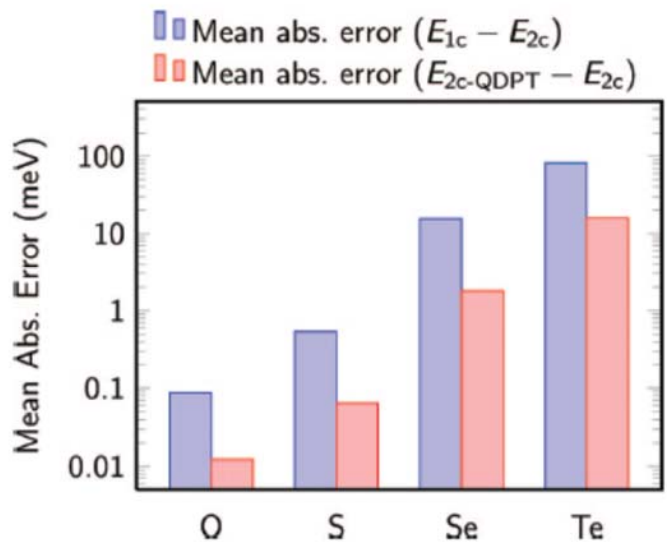

FIG. 6. Mean absolute errors when comparing the 1c (blue) and 2c-QDPT (red) excitation energies with the variational 2c excitation energies. For the relative energies see Table II. Note the logarithmic scale.

Table III presents the number of CSFs in the MRCISD expansions of the calculations presented in Table II. The factors f2c (see Eq. (5)) are also included. The 1c-MRCISD numbers are the sum of the number of singlet and triplet CSFs. The number of CSFs allows for an estimate of the computational cost of the MRCI calculation. Clearly, a 1c calculation is significantly cheaper than a 2c calculation. Since the cost of a 2c-QDPT calculation is dominated by the optimization of the 1c model space wavefunctions, the 2c-QDPT scheme comes essentially at the cost of a 1c calculation. The table also shows that accurate large-scale 2c-MRCISD calculations can be efficiently carried out with Columbus. We note that the 2c-QDPT calculation for Te-acrolein was performed within 2 days on a single CPU core, while the 2c-MRCISD calculation required 1 day using 48 CPU cores on the same computer architecture.

TABLE III. Number of configuration state functions (CSFs) for the MRCISD calculations of Table II and factors by which 2c-MRCISD is more expensive than 1c-MRCISD according to Eq. (5).

| Element Y | 1c-MRCISD | 2c-MRCISD | $f^{2c}$ |
|---|---|---|---|
| O and S | 42 237 550 | 95 153 766 | 2.25 |
| Se | 101 888 896 | 229 817 326 | 2.26 |
| Te | 131 799 096 | 297 305 958 | 2.26 |

**B. LRT-MRAQCC vertical excitation energies**

As a proof of concept, we also conducted vertical excitation calculations on the LRT-MRAQCC level of theory for Te-acrolein. As before, we computed 1c, 2c-QDPT, and variational 2c energies for the singlet ground state and nπ* state as well as for the triplet nπ* and ππ* states. The MRAQCC wavefunctions were based on the same orbitals and reference space as the MRCISD calculations, but only singly excited CSFs were included.

Table IV presents the results for Te-acrolein. It can be noted that the MRAQCC excitation energies are systematically larger than the MRCISD ones due to the neglect of the doubly excited CSFs. Furthermore, at the LRT-MRAQCC level of theory the state with predominant S1 character is not interleaved between the T1 components. However, comparing the 1c and 2c-QDPT energies to the full 2c results reveals the same trends as observed above: while the 1c calculation is off by up to 100 meV, the 2c-QDPT results are within 10 meV (except for the S1, which is off by 28 meV) to the reference energy. The mean absolute error of $\Delta E - \Delta E_{2c}$ is 65 meV for the 1c calculation, but only 7 meV for the 2c-QDPT calculation.

TABLE IV. Energies (eV) of the of the ground state and nπ* singlet and nπ* and ππ* triplet states of Te-acrolein using LRT-MRAQCC including single excitations. Bold digits differ from the 2c-MRCISD reference values.

| State | | 1c | 2c-QDPT | 2c |
|---|---|---|---|---|
| 1 | $S_0$ | 0.0000 | 0.0000 | 0.0000 |
| 2 | $T_1$ | 1.2581 | 1.2302 | 1.2318 |
| 3 | $T_1$ | 1.2581 | 1.2302 | 1.2318 |
| 4 | $T_1$ | 1.2581 | 1.3133 | 1.3062 |
| 5 | $S_1$ | 1.3864 | 1.3406 | 1.3126 |
| 6 | $T_2$ | 1.8120 | 1.8950 | 1.9011 |
| 7 | $T_2$ | 1.8120 | 1.8950 | 1.9019 |
| 8 | $T_2$ | 1.8120 | 1.9129 | 1.9119 |

**C. Size of the model space**

We also investigated the convergence behaviour of the 8 lowest 2c-QDPT energies to the full 2c energies when increasing the size of the QDPT model space. These calculations were only performed for the Te-acrolein, since the SO effects are largest in this system. The mean absolute errors of the excitation energies (analogue to the ones presented in Figure 6) are shown in Figure 7 for CASCI (no external excitations) and MRCIS (only singly external excitations) calculations. As the main goal is the comparison of 2c and 2c-QDPT and since spin polarisation is included already at the MRCIS level, these calculations were performed using MRCIS instead of MRCISD.

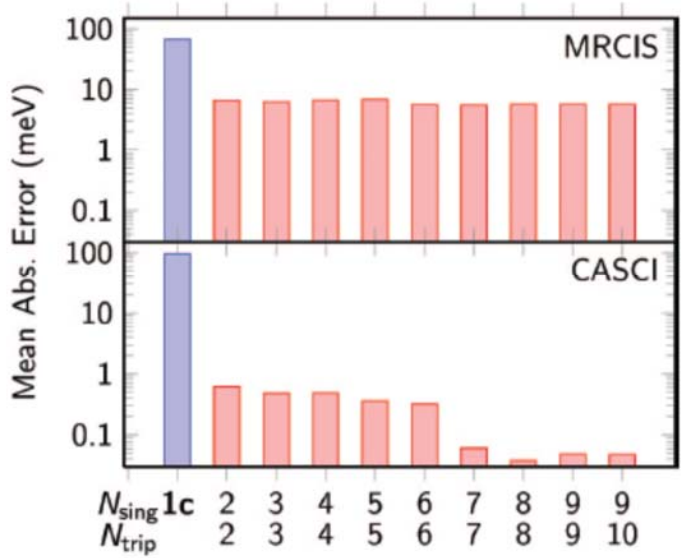

FIG. 7. Mean absolute errors when comparing the 1c (blue) and 2c-QDPT (red) excitation energies with the variational 2c excitation energies for different sizes of the QDPT model space. Note that the size of the model space is Nsing + 3Ntrip.

The left-most values of Figure 7 give the mean absolute errors from the 1c calculations. As in Figure 6, these values are around 100 meV, which is rather large. By going to 2c-QDPT with the minimum model space (2 singlets and 2 triplets give 8 states in 2c formalism) the errors are reduced to 10 meV (MRCIS) or 1 meV (CASCI).

In the case of MRCIS, enlarging the model space beyond the minimum size has almost no effect on the mean errors. Even the largest 2c-QDPT calculation with a model space containing 39 states gives basically the same results as the 2c-QDPT calculation including only 8 states. The 2c wavefunction contains contributions of high-energy singly excited CSFs (spin-polarisation), which are neither present in the 8-state model space nor in the 39-state model space. Recovering these effects in order to reduce the mean absolute errors of the excitation energies would require a significantly larger model space.

On the contrary, in the CASCI case the standard deviations are reduced quickly when increasing the size of the model space. This is because CASCI per se does not include the singly excited CSFs responsible for spin-polarisation. Since the variational 2c wavefunction does not include spin-polarisation, 2c-QDPT does not need to recover it and a fast convergence of 2c-QDPT to variational 2c is obtained.

These findings indicate that for a reasonable 2c-QDPT calculation singly excited CSFs should be included. However, increasing the size of the model space beyond the minimum size incurs significant extra computational cost while not necessarily improving the results.

### D. Potential energy scans along the torsion mode

All calculations presented above were carried out at the ground state equilibrium geometry of the acrolein molecules. In order to observe the effect of SO coupling on the shape of PECs of the excited states, we conducted a rigid scan along the torsion around the C=C double bond (denoted as θ in Figure 5). This internal coordinate was chosen since $1n\pi^*$ and $3\pi\pi^*$ cross along this coordinate.[87] According to the El-Sayed rule[88] SO matrix element between $1n\pi^*$ and $3\pi\pi^*$ should be large, and we expect SO coupling to significantly deform the potentials (at least for the heavy homologues). The scan has been carried out using the method described above, except that the CASSCF calculation state-averaged over 2 singlets and 2 triplets and that MRCIS was employed. For the 2c-QDPT calculation, 2 singlets and 2 triplets were included in the model space.

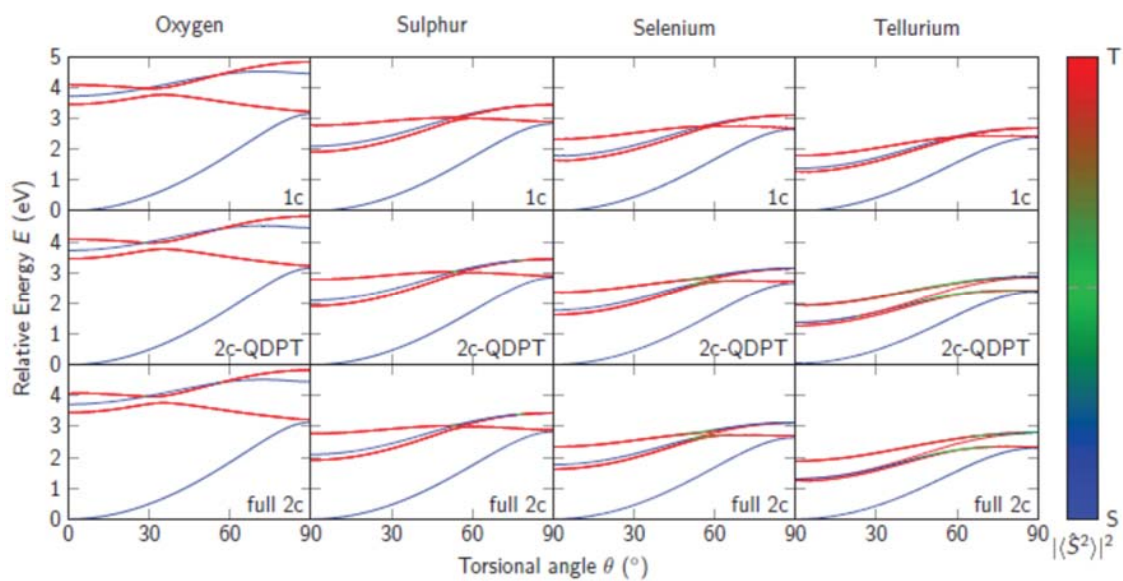

FIG. 8. Potential energy scans of the acrolein systems along the C=C torsion mode. Top row: 1c-MRCIS, middle row: 2c-QDPT-MRCIS, bottom row: 2c-MRCIS. Colors indicate the expectation value of the total spin.

Figure 8 shows the PECs for all acrolein systems based on 1c, 2c-QDPT and 2c energies. The total spin expectation value $\langle \Psi_i | \hat{S} | \Psi_i \rangle$ is indicated using colors, where states that are predominantly singlet are blue and triplets are red. States which arise from a mixture of singlets and triplets appear green.

For all systems, the minima of the ground state and the 1nπ* and 3nπ* states are at θ = 0°, the minimum of the 3ππ* is at θ = 90°. The triplet states 3ππ* and 3nπ* show a weakly avoided crossing between 40° and 60° (depending on Y). The PECs of 1nπ* and 3nπ* are nearly parallel and close to each other. The 1nπ* and 3ππ* states cross as well, as mentioned above.

For O- and S-acrolein, the PECs are basically unaffected by the inclusion of SO coupling via the 2c-QDPT and full 2c methods. For Se- and Te-acrolein, the 2c-QDPT and 2c PECs differ from the 1c PECs. Most notably, the PECs are significantly deformed close to the 1nπ* − 3ππ* crossing. The 2c-QDPT treatment is able to describe these deformations of the PES qualitatively correctly when compared to the full 2c PECs. However, in the Te case the maximum deviations of 2c-QDPT from 2c approach 100 meV and hence the state ordering in the 2c-QDPT calculations is wrong for some torsion angles.

## V. CONCLUSION

A QDPT scheme for the calculation of relativistic two-component wavefunctions has been implemented in the Columbus quantum chemistry program package. The QDPT approach allows to derive two-component wavefunctions based on scalar-relativistic MRCI and MRAQCC wavefunctions. The current implementation allows to use the QDPT method for the treatment of SO coupling in combination with the highly general and efficient MRCI code in Columbus. The 2c-QDPT wavefunctions can be used as good approximations to the variational 2c-MRCI wavefunctions, or as high quality initial guesses for the 2c-MRCISD calculations. Additionally, the compatibility of LRT-MRAQCC and QDPT allows to obtain approximately size-consistent relativistic two-component wavefunctions.

The availability of excited-state gradients and non-adiabatic couplings for the 1c wavefunctions together with the model space Hamiltonian from the QDPT treatment allows to derive 2c gradients and non-adiabatic couplings assuming a slowly varying SO potential. This is interesting, e.g., to perform excited-state dynamics including internal conversion and intersystem crossing on the same footing, e.g., within the Sharc method.89 Additionally, the outlined procedure offers the potential to study SO coupling effects on crossing seams for odd-electron systems with highly correlated wavefunctions at reasonable cost.

The application of 2c-QDPT to the chalcogene analoga of acrolein showed that 2c-QDPT quantitatively agrees with full variational 2c methods for molecules containing first- and second-row atoms. For heavier atoms up to fourth-row, 2c-QDPT agrees qualitatively with the full 2c results. Thus, the more efficient 2c-QDPT approach can be used in place of the more expensive full 2c calculation.

## ACKNOWLEDGMENTS

We gratefully acknowledge support by the Jülich Super Computing Centre through Simlab Molecular Systems and provision of computer time on the general purpose supercomputer cluster JUROPA. S.M., P.M., and L.G. thank the Austrian Science Fond (FWF) for financial support through the Project P25827, the COST Actions CM1204 (XLIC) and CM1305 (ECOSTBio), as well as the Vienna Scientific Cluster (VSC). H.L. acknowledges support by the National Science Foundation (NSF) (Project No. CHE-1213263), the FWF within the framework of the Special Research Program F41, and the Robert A. Welch Foundation (Grant No. D-0005). F.P. is a recipient of a fellowship for postdoctoral researchers by the Alexander von Humboldt Foundation.

Chem. Lett. 3, 3090–3095 (2012).